\documentclass[twocolumn]{jpsj3}

\usepackage{color}
\definecolor{mycolor}{rgb}{1,0,0}

\title{Details of Sample Dependence and Transport Properties of URu$_2$Si$_2$}

\author{Tatsuma D. {\sc Matsuda}$^{1,2}$\thanks{E-mail address:matsuda.tatsuma@jaea.go.jp}, 
Elena {\sc Hassinger}$^2$, Dai {\sc Aoki}$^2$, Valentin {\sc Taufour}$^2$, \\
Georg {\sc Knebel}$^2$, Naoyuki {\sc Tateiwa}$^1$, Etsuji {\sc Yamamoto}$^1$, Yoshinori {\sc Haga}$^1$,\\
 Yoshichika {\sc \=Onuki}$^{1,3}$, Zachary Fisk$^{1,4}$ and Jacques {\sc Flouquet}$^2$ 
}
\inst{$^1$ ASRC, JAEA, Tokai, Ibaraki 319-1195, Japan\\
$^2$ INAC/SPSMS, CEA-Grenoble, 38054 Grenoble, France\\
$^3$ Graduate School of Science, Osaka University, Toyonaka, Osaka 560-0043, Japan\\ 
$^4$ University of California, Irvine, Calfornia 92697, U.S.A.
}
\abst{%
Resistivity and specific heat measurements were performed in the low carrier unconventional superconductor URu$_2$Si$_2$ on various samples with very different qualities.
The superconducting transition temperature ($T_{\rm SC}$) and the hidden order transition temperature ($T_{\rm HO}$) of these crystals
were evaluated as a function of the residual resistivity ratio (RRR).
In high quality single crystals the resistivity does not seem to follow a $T^2$ dependence above $T_{\rm SC}$ ,
indicating that the Fermi liquid regime is restricted to low temperatures. 
However, an analysis of the isothermal longitudinal magnetoresistivity points out that the $T^2$ dependence may be ``spoiled" by  residual inhomogeneous superconducting contribution.
We discuss a possible scenario concerning the distribution of $T_{\rm SC}$ related with 
the fact that the hidden order phase is very sensitive to the pressure inhomogeneity.
}

\kword{resistivity, specific heat, superconductivity, hidden order, URu$_2$Si$_2$}

\begin{document}
\maketitle

\section{Introduction} 

The rush to elucidate the enigma of the nature of the hidden order (HO) state of URu$_2$Si$_2$ which appears below 
$T_{\rm HO}\sim$17.5 K in URu$_2$Si$_2$\cite{Palstra1,1986Schlabitz,1986Maple}has led to a large variety of macroscopic as well as microscopic experiments.\cite{2007Amitsuka}
However, no detailed report has been given on the link between crystal growth, 
characterization of the crystal purity, and their influence on physical properties of URu$_2$Si$_2$. 
In the present article we discuss specific heat and transport measurements on various single crystals, with the focus on the anomalies at the HO  
and the superconducting (SC) transition at $T_{\rm HO} \sim 17.5$~K and $T_{\rm SC}\sim 1.2$~K as a function of the residual resistivity ratio (RRR) of the resistivity between room temperature and $T=2$~K.
The aim is to give a sound basis for future experimental studies which require a deep knowledge of the material.  
We find that the physical properties depend on the sample purity (given by the RRR) as well as the position of the crystals with respect to the crystal's growth direction and its radial distance from the center of the crystals.  

The apparent consensus on the high pressure phase diagram of URu$_2$Si$_2$ is that at low pressure ($P$) 
the ground state is the exotic HO phase but above a critical pressure $P_x \sim$ 0.5 GPa it switches to a conventional antiferromagnetic phase (AF).\cite{2007Amitsuka,1999Amitsuka,2008Hassinger,2010Butch} 
Now, it is established that bulk superconductivity is associated with the HO while it is suppressed in the AF state \cite{2006Sato, 2008Hassinger}. 
The persistence of a tiny ordered moment below $P_x$ or the observation of a superconducting transition in resistivity up to $P_{\rm SC} > 2 P_x$ emphasizes that the separation between intrinsic and extrinsic phenomena is still not fully solved. 
Due to the high sensitivity to pressure and uniaxial stress (presumably associated with the switch from HO to AF in only a few kbars), 
pressure gradients near stacking fault or other defects can easily produce local deformations which will stabilize AF inside the HO and SC inside the AF states. 
Furthermore, the possibility of correlated defects with an unusual nanostructure network is an open question. 

\begin{table}[tbh]
\caption{Atomic coordinates and thermal parameters of URu$_2$Si$_2$ at 300~K from x-ray measurements. 
$R$ and $wR$ are the reliability factors and $B$ is the equivalent isotropic atomic displacement parameter.}
\label{table1}
\begin{center}
\begin{tabular}{ lcr l l l} \hline 
						& atom	&\multicolumn{3}{l}{\ \ position}																				&\multicolumn{1}{c}{$B$\ \ \ }\\ 
$I4/mmm$($\sharp$139)	& 	 (site)					&\multicolumn{1}{l}{$x$} 		& \multicolumn{1}{l}{\ $y$}& \multicolumn{1}{l}{$z$}	&\\
\hline 
$a=$ 4.1327(3)			&	U (2$a$)		&0\ 	&\ 0	&0			&	\ 0.59(3) 	\\
$c=$ 9.5839(6)			&  Ru(4$d$)	 	&0\ 	&1/2	&1/4			&	\ 0.47(3)	\\
$V=$ 163.685(17)	&  Si (4$e$)		&0\ 	&\ 0	&0.3724(5) 	&	\ 0.49(7)		\\
																			
 \multicolumn{2}{c}{($R$= 2.23, $wR$= 4.89)  	}																	\\
\hline
\end{tabular}
\end{center}
\end{table}

\section{Experimental} 

Single crystals of URu$_2$Si$_2$ were grown using the Czochralski method in a tetra-arc furnace under argon gas atmosphere, 
both at CEA-Grenoble and at JAEA in Tokai.
Here, the purity of the usual uranium metal is typically 99.9 - 99.95 \%. 
As the purity of the uranium metal is lower than that for Ru and Si materials, 
the quality of the URu$_2$Si$_2$ sample seriously depends on the purity of uranium itself. 
We thus purified a uranium ingot using the solid state electro-transport (SSE) method under ultrahigh vacuum, 
which was found to be extremely effective in removing impurities such as Fe and Cu in the uranium ingot.\cite{SSE} 
A URu$_2$Si$_2$ ingot was also subsequently annealed using the SSE method under ultrahigh vacuum as reported in ref. \citen{2008matsuda}.  
As will be shown below, the crystals with the highest RRR were not obtained in the central core of the crystals but near the free surface.

Most characterizations were made using a Quantum Design PPMS apparatus with magnetic fields up to 9 T and temperatures down to 0.4 K via 
measurements of magnetoresistivity and specific heat.  
Further experiments down to 80 mK used homemade dilution refrigerators. 
Special attention is paid to the different field and current directions in $T$ and $H$ dependent measurements with respect to the $c$ axis. Mainly we measured the longitudinal magnetoresistivity in order to minimize the orbital effect.   
The resistivity measurements were carried out by standard four point contacts AC method. 
For the low temperature measurements down to 0.1 K, the maximal current was $I$ = 50 $\mu$A. 
The measurement frequency of $\sim$17.54 Hz was chosen by considering bandpass and empirical condition of our measurement system.
The signal was amplified by a transformer and a pre-amplifier by factor of 10$^5$ and detected with a lock-in amplifier.

X-ray measurements were performed by a Rigaku RAXIS RAPID imaging plate area detector with graphite monochromated Mo-K$\alpha$ radiation. 
We selected a small high quality single crystal with dimensions of 0.05$\times$0.05$\times$0.01\,mm$^3$ in order to minimize the
absorption as well as the secondary extinction effects, which generally make the structural determination difficult. 
The single crystal sample of URu$_2$Si$_2$ were mounted on a glass fiber. 
The crystal structure data at room temperature are summarized in Table~\ref{table1}. 
The lattice parameter and the $z$ parameter of Si site are basically in good agreement with a reported data previously.\cite{1985structure}

\section{Results} 

\subsection{Phase transitions: HO and SC}
\begin{figure}[h]
\begin{center}
\includegraphics[width=1 \hsize]{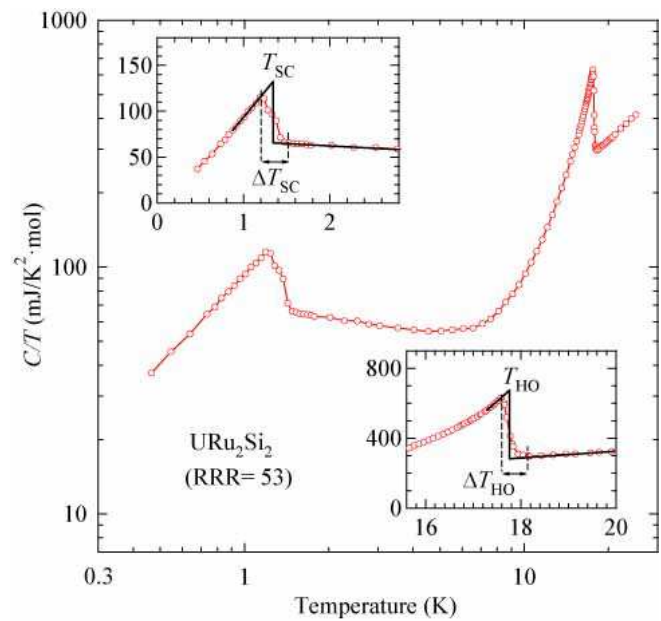}
\caption{(Color online) Temperature dependence of the specific heat of URu$_2$Si$_2$ using a sample with the residual resistivity ratio (RRR) equal to 53. 
The inserts show the definition of the transition width of the superconducting and HO transition, respectively.}
\label{fig:CT}
\end{center}
\end{figure}

\begin{figure}[h]
\begin{center}
\includegraphics[width=1\hsize]{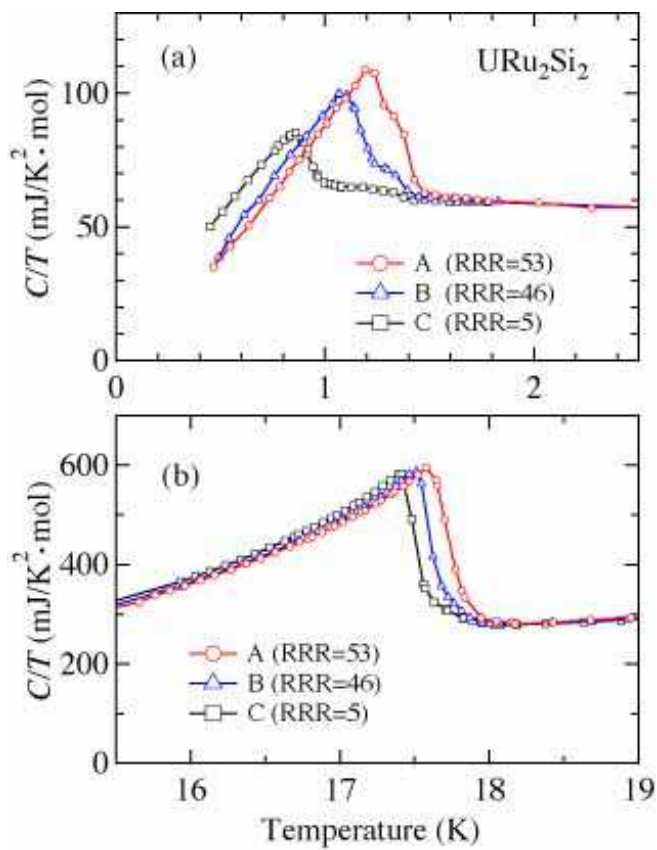}
\caption{(Color online) Temperature dependence of the specific heat for the temperature range around (a) superconducting and 
(b) hidden order temperature for three samples with different RRR.}
\label{fig:C_order4}
\end{center}
\end{figure}

\begin{figure}[h]
\begin{center}
\includegraphics[width=1 \hsize]{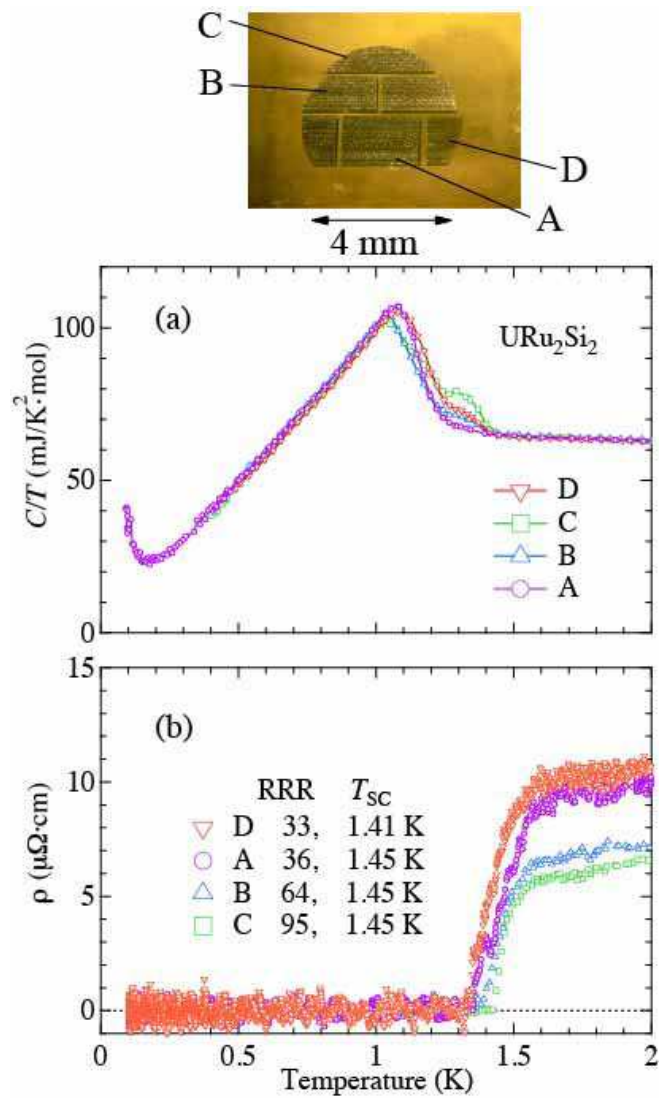}
\caption{(Color online) Temperature dependence of (a) specific heat divided by temperature and 
(b) resistivity measured on different quality samples with RRR more than 30 cut from the same disk of a single crystal as shown in the upper picture. The $T_{\rm SC}$ indicated in the lower panel indicated the midpoint of the transition. }
\label{fig:Cp_resist}
\end{center}
\end{figure}

In order to determine the influence of sample quality on the bulk transition temperature of SC and HO, 
we performed specific heat measurements on crystals with different RRR$ =\rho$(300~K)/$\rho $(2~K) values. 
Figures~\ref{fig:CT}, \ref{fig:C_order4},  and \ref{fig:Cp_resist} show the temperature dependence of the specific heat $C$ divided by temperature for various samples.  Sharp specific heat anomalies occur at $T_{\rm SC}$ and $T_{\rm H0}$. 
The clear mark of the phase transition is linked to  incomplete and complete gap opening at $T_{\rm H0}$ and $T_{\rm SC}$, respectively.

\begin{figure}[h]
\begin{center}
\includegraphics[width=1 \hsize]{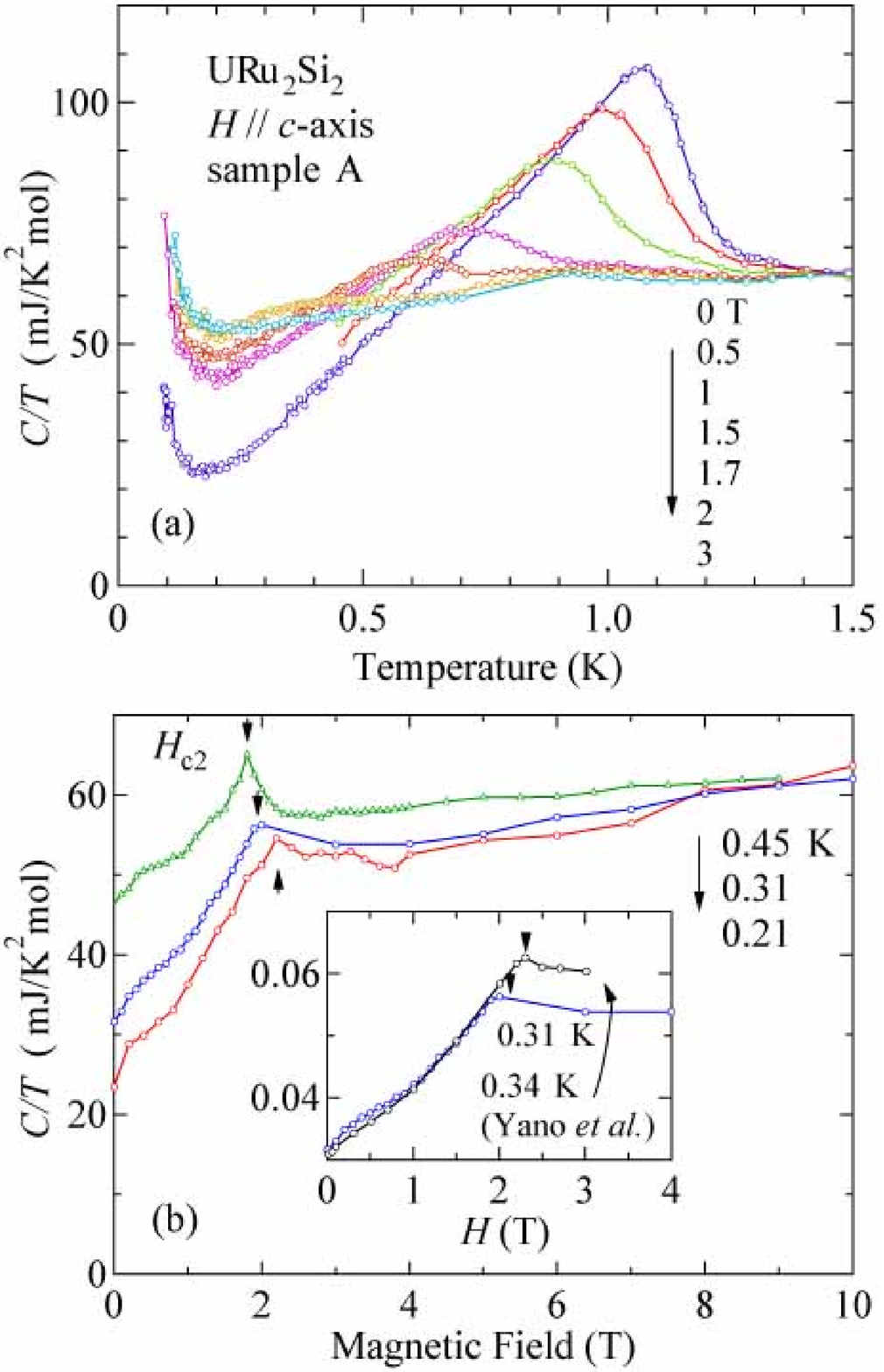}
\caption{(Color online) Temperature dependence of specific heat under various magnetic fields (a) and field dependence of specific heat $C/T$ at low temperatures (b). 
The inset in the lower panel shows the low field range in an enlarged scale in comparison to the field dependence reported in ref.~\citen{2008Yano}.} 
\label{fig:Cp_Hdep}
\end{center}
\end{figure}

We first look at the dependence of $T_{\rm H0}$ and $T_{\rm SC}$ as a function of the residual resistivity RRR  and in the transition width as defined in the insets of Fig.~\ref{fig:CT}. 
Figures~\ref{fig:C_order4}(a) and \ref{fig:C_order4}(b) show the temperature dependence of $C/T$ around SC and HO transition temperatures 
for three samples of different quality determined by the residual resistivity ratio (RRR) at 2 K  for current along $a$-axis.
The strong sample dependence of the shape of $C/T$ at SC transition down to rather low RRR value samples confirms previous reports. \cite{1991Rami,1991Hassel}
Furthermore, a quite similar variation is observed for the HO transition temperature in dependence of the sample quality as shown in Fig. \ref{fig:C_order4}(b).  
The variation of $T_{\rm SC}$ and $T_{\rm HO}$ going from RRR  = 53 to 5 is about 0.5 K for $T_{SC}$ and for the hidden order transition $T_{\rm HO}$ about 0.2~K. 
For both the SC and HO transitions the transition temperatures shift to lower temperatures with decreasing RRR.

To clarify this feature,  we cut a disk out of a single crystal and then cut that disk into different samples, see Fig.~\ref{fig:Cp_resist}. 
This allows making the link between RRR value and crystal location in the ingot.  
All selected samples have a RRR $> 30$, the highest RRR $= 95$ is observed for the most outer sample with a large free surface and not in the center of the disk. 
It is remarkable that below the temperature where $C/T$ reaches its low temperature maximum as well as above $1.6\,{\rm K}$ the specific heat of all samples almost coincides and differences in the specific heat appear only at the transition. 
$T_{\rm SC}$ determined from specific heat varies between $1.15 < T_{\rm SC} < 1.2$~K and is significantly lower than the $T_{\rm SC}$ determined from the resistivity. 
Most remarkable, even the sample highest RRR shows the most pronounced the double transition in specific heat. This observation seems to be against the arguments that for perfect crystals, an unique SC phase transition will occur.\cite{1991Rami,1991Hassel}
Clearly increasing RRR does not lead to obtain an unique specific heat anomaly at $T_{\rm SC}$. 

Figure~\ref{fig:Cp_Hdep} (a) shows $C/T$ for different magnetic fields and Figure~\ref{fig:Cp_Hdep} (b) the field variation of $C/T$ at constant temperature  
for sample A, which has nearly a single jump in the specific heat at $T_{\rm SC}$, whereas the width  $\Delta T_{\rm SC} \approx 0.2$~K is still rather large.  
In the field sweep the location of the upper critical field $H_{\rm c2}$ can be clearly seen. 
However, the high quality of the material is demonstrated by the collapsing value of the 
extrapolated term $\gamma$ of the linear temperature dependence of $C/T$ at $T \to 0\,{\rm K}$. The increase of $C/T$ below 150~mK is due to  the supplementary nuclear hyperfine contribution.

The field dependence of $C/T$ shows the linear increase with the slight convex curvature at low fields.
The results at $0.3\,{\rm K}$ are consistent with previous reports~\cite{2008Yano,1990Fisher}, 
although the values of $H_{\rm c2}$ and $C/T$ above $H_{\rm c2}$ are slightly shifted to lower values,
probably due to the hyperfine contribution related to impurities, to the difference of $T_{\rm SC}$, 
and to the presence of a single or double transition in the specific heat.

\begin{figure}[h]
\begin{center}
\includegraphics[width=1 \hsize]{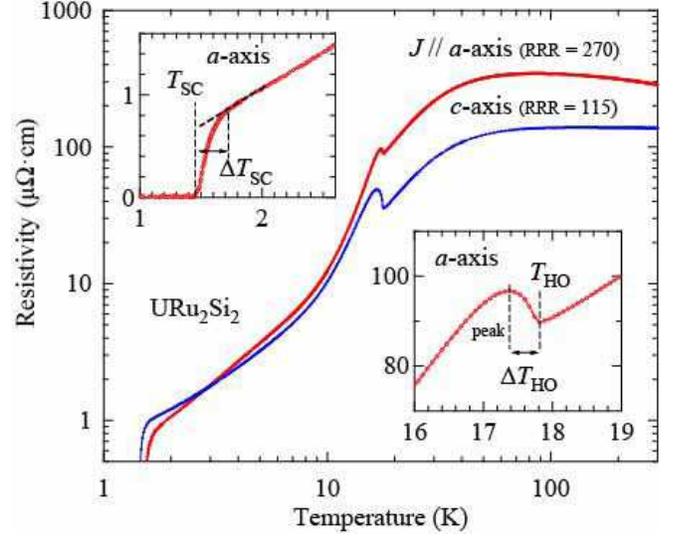}
\caption{(Color online) Temperature dependence of the electrical resistivity of the highest quality sample for current along $a$- and $c$-axis. 
The insets show the definition of $T_{\rm SC}$ and $T_{\rm HO}$ and the width of the transitions, respectively. } 
\label{fig:rhodef}
\end{center}
\end{figure}

Figure \ref{fig:rhodef} shows the temperature dependence of the resistivity of the best studied single crystal with a RRR of 270 for the current $J \parallel a$ 
and RRR = 115 for $J \parallel c$ on a double-logarithmic scale. 
The inset shows the determination of $T_{\rm SC}$, $T_{\rm HO}$ and the corresponding widths  $\Delta T_{\rm SC}$ and  $\Delta T_{\rm HO}$, respectively. 
Generally, the RRR is larger for $J \parallel a$, while the anomaly at $T_{\rm HO}$  is more pronounced and the transition width at $T_{\rm SC}$ smaller for $J \parallel c $.  

\begin{figure*}[t]
\begin{center}
\includegraphics[width=15cm]{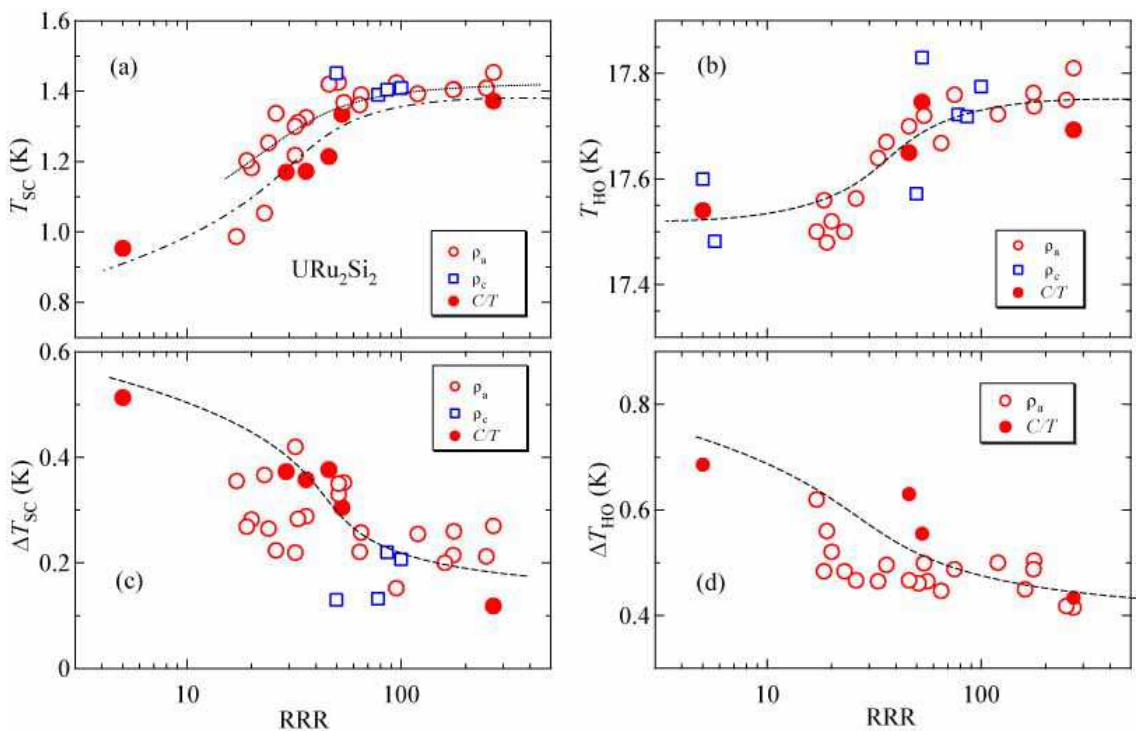}
\caption{(Color online) Dependence of the transition temperatures ((a)$T_{\rm SC}$ and (b)$T_{\rm HO}$) and (c)(d) the width of those transitions of the sample quality defined by the residual resistivity ratio RRR at $T=2$~K. }
\label{fig:RRR_TSC}
\end{center}
\end{figure*}

In order to clarify the relation between sample quality and the superconducting transition temperature $T_{\rm SC}$  quantitatively,   
$T_{\rm SC}$ and the transition width $\Delta T_{\rm SC}$ are plotted as a function of RRR in Fig.~\ref{fig:RRR_TSC} (a) and (c), respectively.
Here, $T_{\rm SC}^{C}$ is defined by the entropy balance in specific heat anomaly and $T_{\rm SC}^{\rho}$ by zero resistivity, respectively. 
The transition width $\Delta T_{\rm SC}$ is defined by the difference between the onset and peak position in the specific heat and the width of the resistive transition as shown in the inset of Fig.~\ref{fig:rhodef}. 

As already mentioned above, due to the non-quadratic temperature dependence of the resistivity at low temperatures just above the superconducting transition, 
the extrapolation to $T = 0$ K is ambiguous and therefore, 
we take the RRR at $T = 2$~K above the onset of the superconducting transition. Figure~\ref{fig:RRR_TSC} (a) clearly indicates the saturation tendency of $T_{\rm SC}$ with increasing RRR. 
Above RRR $\sim$ 50, $T^{\rho}_{\rm SC}$ reaches $\sim$ 1.4~K while $T^{C}_{\rm SC}$ is slightly lower. 
Remarkably, the transition width becomes narrower above RRR $\sim 50$. 
This tendency is observed for both the specific heat and resistivity measurements.  
In Figs.~\ref{fig:RRR_TSC} (b) and (d), $T_{\rm HO}$ and $\Delta T_{\rm HO}$ are plotted as a function of RRR. 
With increasing sample quality, $T_{\rm HO}$ is increasing and shows the same saturation tendency above RRR $\sim$ 50.  

\subsection{Resistivity in the Hidden Order}
Having looked at the SC and HO anomalies at $T_{\rm SC}$ and $T_{\rm HO}$ as a function of RRR, 
let us now focus on the temperature dependence of the resistivity below $T_{\rm HO}$.
At low temperatures, the resistivity does not follow a $T^2$ behavior over a large temperature range just above the superconducting transition. 
This has been already pointed out in refs.~\citen{McE,1993Shumidt,1986Maple,Kamran} and confirmed in and recent works of refs.~\citen{2008Hassinger,2010Elena,2011Hassinger,2011Tateiwa,2011Tateiwa2}.

\begin{figure}[h]
\begin{center}
\includegraphics[width=1 \hsize]{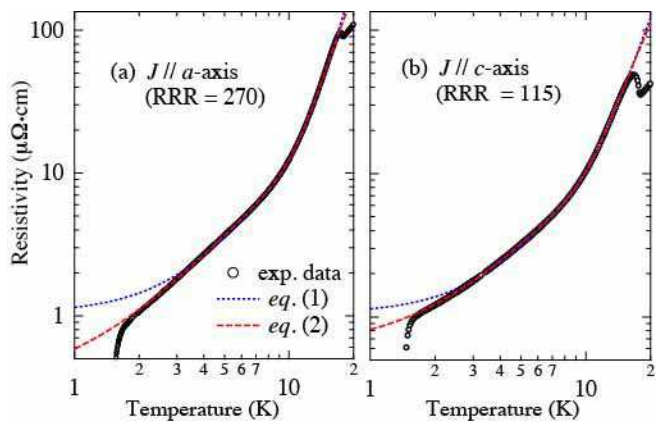}
\caption{(Color online) Low temperature behavior of $\rho$ in the best sample for current along (a) $a$- and (b) $c$-axis. Broken lines represent fitting lines  described in the text.}
\label{fig:fit}
\end{center}
\end{figure}

At the hidden order transition $T_{\rm HO}$, the resistivity shows a jump due to the partial gapping by the nesting of FS with a loss of carriers. 
In the HO state, the temperature dependence of the resistivity has been analyzed often using eq.~(\ref{eqn1}) with gapped spin waves in an antiferromagnet,\cite{spinAF}  
 but there is no justification to apply it to the HO and to the AF phase characterized by only Ising type fluctuations. 
Such a fit does give an order of magnitude of the $A$ coefficient of the $T^2$ term and of the gap opened at the Fermi surface reconstruction at $T_{\rm HO}$. 
\begin{equation}
\label{eqn1}
\rho(T) = \rho_0 + AT^2 + BT(1+\frac{2T}{\Delta})\exp(-\frac{\Delta}{T})
\end{equation} 
\begin{table}[thb]
\caption{Fitting parameters using formula~(\ref{eqn1}) for the temperature dependence of the resistivity of a single crystal with a RRR = 270, see figure \ref{fig:fit}.}
\label{tab}
\begin{center}
\begin{tabular}{ c ccc cc  } \hline 
		& 	$\rho_0$		&	$A$&	$B$	&	$\Delta$ 	\\ 
	& ($\mu\Omega\cdot$cm)		&	($\mu\Omega\cdot$cmK$^{-2}$)	&	($\mu\Omega\cdot$cmK$^{-1}$)	&	(K)	\\ 
\hline 
$J\| a$-axis \\
	(2-15K)	&	1.05				&	0.099		&	259		&		76.4	\\
\hline
$J\| c$-axis \\
	(2-15K)	&	1.06				&	0.081		&	72.6		&		65.4	\\
\hline
\end{tabular}
\end{center}


\caption{Fitting parameters using formula 2.}
\label{table2}
\begin{center}
\begin{tabular}{ c cccc c } \hline 
			& 	$\rho_0$	&	$x$	&	$A$	&	$B$	&	$\Delta$ 	\\ 
& 	($\mu\Omega\cdot$cm)	&		&	($\mu\Omega\cdot$cmK$^{-x}$)	&	($\mu\Omega\cdot$cm)	&	(K)	\\ 
\hline 
$J\| a$-axis \\
	(2-14K)	&  0.32		& 	1.58	&	0.26		&	10800		&		85	\\
\hline
$J\| c$-axis \\
	(2-14K)	&  0.61		& 	1.58	&	0.21	&	4000		&		76	\\
\hline
\end{tabular}
\end{center}
\end{table}
As pointed out above, at low temperatures the resistivity deviates significantly from a Fermi liquid $AT^2$ behavior. 
The best fitting parameters by the formula~(\ref{eqn1}) from 2 to 15 K are summarized in Table II. 
However, as shown in Fig.~\ref{fig:fit} the fit deviates significantly from the experimental data.
Considering the deviation from $AT^2$ at low temperatures and the partial gap opening at $T_{\rm HO}$, a better fit to the experimental data is given using a simple formula: 
\begin{equation}
\label{eqn2}
\rho(T) = \rho_0 + AT^x + B\exp(-\frac{\Delta}{T}).
\end{equation}
This model well describes over a  wide  range the temperature dependence of the resistivity below $T_{\rm HO}$. 
The best fitting parameters by the formula~(\ref{eqn2}) from 2 to 14 K are summarized in Table III.
The important point is that the value of exponent $x$ deviates from 2, again. 
The comparison between two different fits over the same temperature interval are shown in Fig.~\ref{fig:fit} and clearly, a better matching is achieved with eq.~(\ref{eqn2}).

\begin{figure}[h]
\begin{center}
\includegraphics[width=0.9 \hsize]{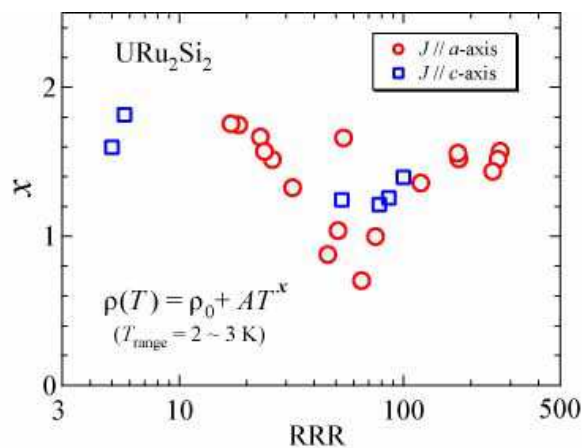}
\caption{(Color online) Sample dependence of the power $x$ of $\rho$($T$) at low temperatures .}
\label{fig:PowX}
\end{center}
\end{figure}
Figure~\ref{fig:PowX} shows the power $x$ of $\rho$(T) at low temperatures above the superconducting transition as function of the RRR.  
This plot obviously indicates the small value of $x$ for both current directions compared to $x = 2$ expected for a Fermi liquid, and $x$ seems to be close to 1.5 for large RRR samples, independent of the current direction. 
On the other hand, for the low RRR  samples, the impurity scattering contribution, which would follow a $T^2$ behavior, should be larger. 
This may be the reason that the $x$ value becomes close to 2 for samples with small RRR.

\begin{figure}[h]
\begin{center}
\includegraphics[width=1 \hsize]{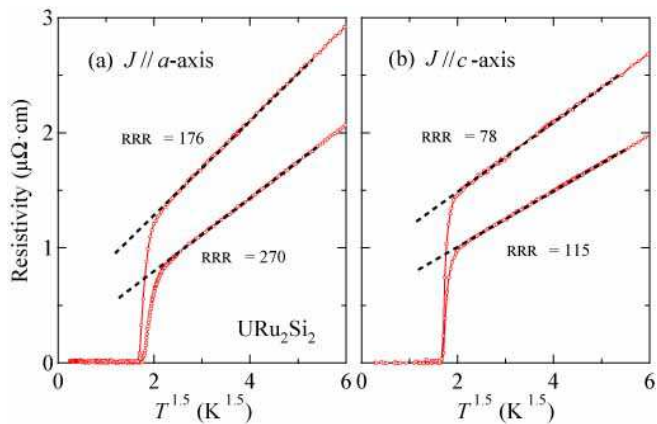}
\caption{(Color online) Temperature dependence of electrical resistivity as a function of $T^{1.5}$ for the best and 2nd samples.}
\label{fig:rho_T15}
\end{center}
\end{figure}

Although the determination of an exact value for the anisotropy of $x$ is difficult to get from our results, 
we plot the temperature dependence of resistivity at low temperature as a function of $T^{1.5}$ in Fig.~\ref{fig:rho_T15} for two very high quality samples for current along $a$- and $c$-axis, respectively. 
At least below 3 K, $x =1.5$ seems to be good expression for the temperature dependence of resistivity for both directions.  
It should be noted that the linear dependence of resistivity was which reported in ref.~\citen{Kamran} for $J\| a$ in a RRR$\sim50$ sample taken from same ingot 
appears to correspond to a ``crossover'' regime where tiny differences in crystal growth lead to large difference in $x$.
The power $x$ of the resistivity of samples cut from the same ingot with roughly the same RRR at low temperatures can vary from $x\sim 1$ to $x\sim1.5$. 
Thus it seems difficult to get a final statement on the detailed temperature dependence of the resistivity.

As discussed above, the resistivity in a limited temperature region shows non-Fermi liquid behavior. 
This is a very unusual behavior for three dimensional metal systems if the electronic bandwidth (BW) is large enough. 
The difficulty to reach a $T^2$ regime can be associated with the fact that, at $T_{\rm HO}$, due to FS reconstruction, a compensated semimetal is 
built with very small bandwidth (5 K) for certain orbits. 
This weakness is directly related to strong renormalization of the band mass and the correlated mass due to the interaction between the quasiparticles. 
Evidence of such a weakness was recently seen in the large changes of Fermi surfaces
when a high magnetic polarization is induced by magnetic field applied along the easy magnetization axis.\cite{2009Shishido,2011Altarawneh,2011Malone}
This leads to the observation of a Fermi liquid $T^2$ regime only below $\mbox{BW}/k_{\rm B} \sim 0.5\,{\rm K}$.
This statement is supported by the smooth increase of $C/T$ and the large increase of the electronic Gr\"{u}neisen parameter on cooling.\cite{1990Ohkawa,Flouquet}

\begin{figure}[th]
\begin{center}
\includegraphics[width=0.9 \hsize]{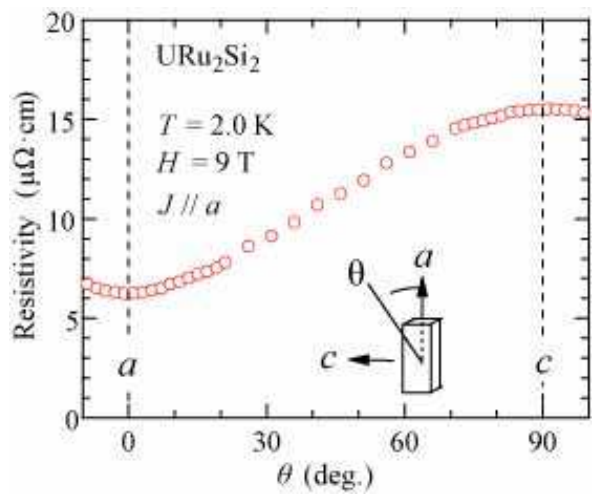}
\caption{(Color online) Angular dependence of magnetoresistance for a middle quality sample (RRR $= 51$).}
\label{fig:angle}
\end{center}
\end{figure}

A possible reason for the lack of observing the Fermi liquid law might be the presence of a parallel scattering channel in the resistivity.
This may happen if SC droplets or filaments exist in the materials.
Such a ``parasitic'' channels may become more efficient as RRR increases,
since the long electronic mean free path may allow to transfer informations
between intrinsic and extrinsic components.
By applying a magnetic field, one may wipeout effects with SC origin.
This possibility prompted the measurement of the field dependence of the longitudinal magnetoresistivity 
for both directions  at different temperatures.
In these experiments the longitudinal configuration is necessary in order to avoid the contribution of
the transverse magnetoresistivity which is sensitive to orbital effects and
thus to $\omega_{\rm c}\tau$ ($\omega_{\rm c}$: cyclotron frequency, $\tau$: scattering lifetime). 
Excellent adjustments between field direction and current direction are necessary to suppress the orbital effect.
Figure~\ref{fig:angle} shows the angular dependence of the magnetoresistivity at $2\,{\rm K}$ and at $9\,{\rm T}$.
The fine tuning of the orientation was realized by slightly tilting the field angle.
Let us notice the huge difference between longitudinal and transverse magnetoresistance
in agreement with the classification of URu$_2$Si$_2$ as a compensated semimetal 
with carrier concentration near $0.06$ per U atom.

\begin{figure}[th]
\begin{center}
\includegraphics[width= 0.9 \hsize]{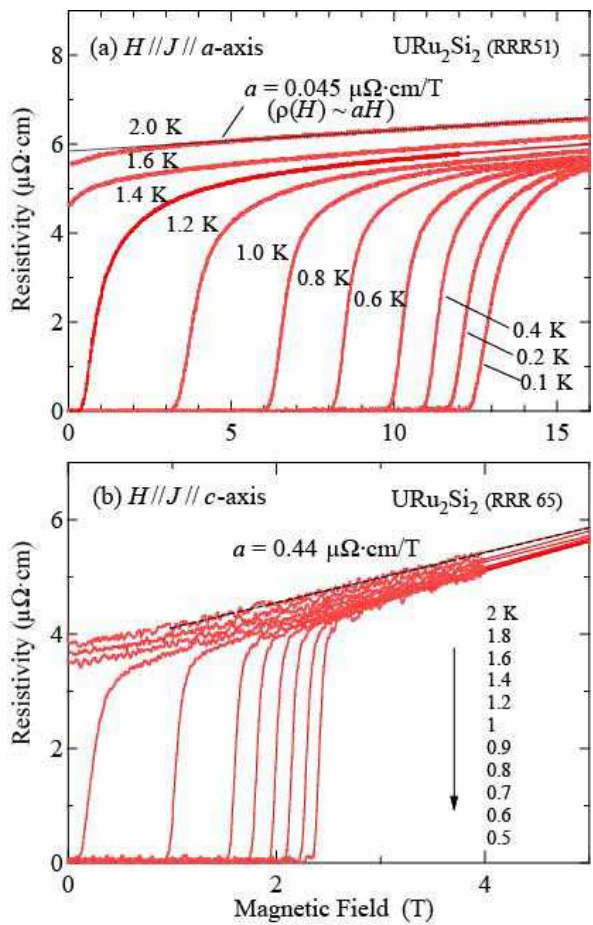}
\caption{(Color online) Longitudinal magnetoresistance ($J\| H\| a$- and $c$-axis) for  middle quality samples (RRR $=51$ and 65, respectively) of URu$_2$Si$_2$.}
\label{fig:18T}
\end{center}
\end{figure}

\begin{figure}[h]
\begin{center}
\vspace{0.5cm}
\includegraphics[width=1 \hsize]{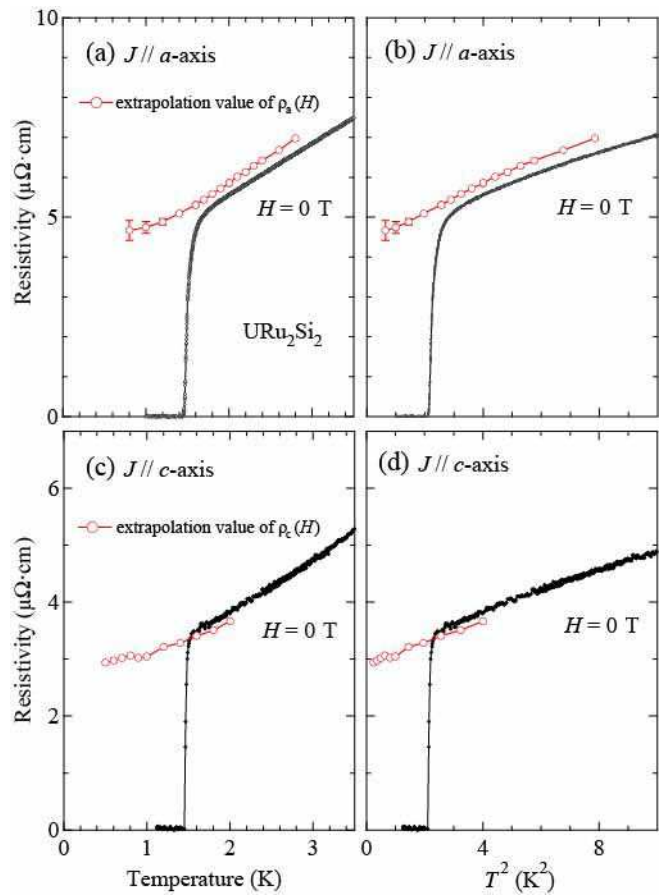}
\caption{(Color online) The extrapolation values of (a) $\rho_a(H\rightarrow 0)$ and (c) $\rho_c(H\rightarrow 0)$ from the normal state at low temperatures and 
(b)(d) $T^2$ plot of those, respectively.}
\label{rho_T_ext}
\end{center}
\end{figure}

The striking feature in Figs.~\ref{fig:18T} (a) and (b) is that the resistivity can be described with two terms,
\begin{equation}
\rho (T,H) = \rho_0 (H)  + \rho_{\rm e} (T,H=0),
\end{equation}
where $\rho_0 (H)$ is a temperature independent linear-$H$ term and 
$\rho_{\rm e} (T,H=0)$ represents the extrapolation of the linear $H$ term at $H=0$.
The coefficients for $a$ and $c$-axis are equal to
$a_{a\mathchar`-\rm{axis}}=0.045\,\mu\Omega\!\cdot\!{\rm cm}/{\rm T}$ and
$a_{c\mathchar`-\rm{axis}}=0.44\,\mu\Omega\!\cdot\!{\rm cm}/{\rm T}$, respectively.
The residual linear $H$ dependence of the magnetoresistivity seems to be related with
the suggestion of ref.~\citen{1990Ohkawa}
that $H$ scan reveals the impurity Kondo distribution which occurs in the lattice.
Furthermore differences between $a_{a-\rm{axis}}$ and $a_{c-\rm{axis}}$ appears 
roughly linked to the size of the $H$ induced magnetic polarization,
which is given by the value of the magnetic susceptibility $\chi_{c\mathchar`-{\rm axis}} > 5 \chi_{a\mathchar`-{\rm axis}}$.
The $A$ coefficients obtained for $a$ and $c$-axis are 
$A_{a\mathchar`-{\rm axis}}=0.28\,\mu\Omega\!\cdot\!{\rm cm K^{-2}}$ and
$A_{c\mathchar`-{\rm axis}}=0.20\,\mu\Omega\!\cdot\!{\rm cm K^{-2}}$, respectively,
as shown form the $T$ dependence of $\rho_{\rm e}$ (Fig.~\ref{rho_T_ext}).

\begin{figure}[h]
\begin{center}
\includegraphics[width=0.9 \hsize]{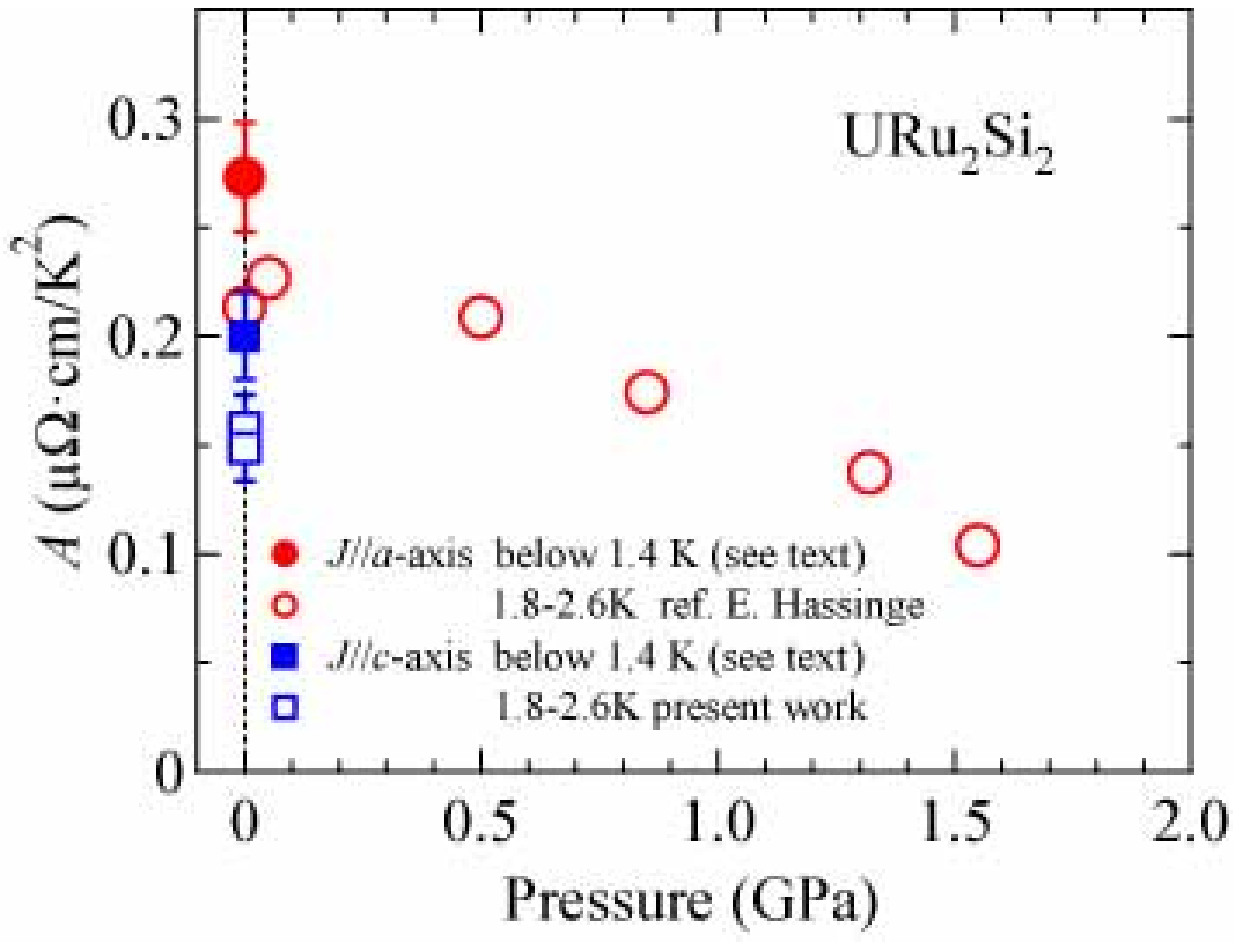}
\caption{(Color online) The $A$ coefficient of the Fermi liquid behavior of the resistivity with data under pressures (cited from \citen{2008Hassinger}).
Extrapolated value from longitudinal magnetoresistivity is indicated by full symbols.}
\label{fig:A_coff}
\end{center}
\end{figure}

In Fig.~\ref{fig:A_coff} the $A$ value obtained by a $T^2$ fit of the data for $J \parallel a$ and $c$-axis
from $1.8\,{\rm K}$ to $2.6\,{\rm K}$ are represented 
and compared at $P=0$ with the previous extrapolated values obtained through the $T$
dependence of the longitudinal magnetoresistivity.
For $J\parallel a$-axis, measurements were recently reported in the AF phase above $P_x$.
Using the measured electronic Gr\"{u}neisen parameter of URu$_2$Si$_2$
the compressibility of $0.5 \times 10^{-6}\,{\rm bar}^{-1}$, 
excellent extrapolation was obtained at $P=0$.
Thus evidences are given that, at $P = 0$, the entrance in the Fermi liquid will occur just below $T_{\rm SC}$
and that SC phase transition reveals an unusual feature.


\section{Discussion} 
\begin{figure}[h]
\begin{center}
\includegraphics[width= 0.9 \hsize]{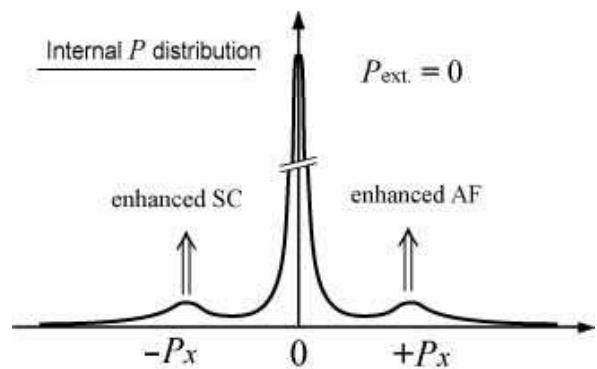}
\caption{Schematic image of internal pressure distribution in URu$_2$Si$_2$.
From the value of residual tiny moment, the tails will represent only few percent of the major sharp distribution.}
\label{fig:P1}
\end{center}
\end{figure}

The link between the intrinsic and extrinsic properties of URu$_2$Si$_2$ crystals 
and their purity as defined by RRR but also by their position in the crystal boule remains complex.
Improving the crystal quality as detected by an increase of RRR 
enhances the electronic mean free path and thus may facilitate the coupling between 
``parasitic extrinsic'' domains such as AF or SC droplets or other exotic nanostructure arrangements.
The simple image (Fig.~\ref{fig:P1}) is that of an enhanced SC component favored by a 
negative internal pressure gradient and of a surviving residual AF volume generated by a positive pressure gradient.
Thus the picture will be a main sharp distribution centered at $P=0$ with two broad tails 
at positive and negative pressure shifted roughly by $\pm P_x$ with a broad width comparable to $P_x$.
This frame may explain why SC detected by resistivity survives often up to $2P_x$.
However, recent Larmor nuclear precession experiments~\cite{2010Niklowitz,2011Bourdarot} indicate that
the a variation of the $c/a = \eta$ ratio of the lattice parameters results to a the lattice inhomogeneity and corresponds to a Gaussian distribution near with a full width half-maximum around $\approx 5\times 10^{-4}$.
Assuming a compressibility of $0.5\times 10^{-6}\,{\rm bar}^{-1}$,
the pressure inhomogeneity will be around $1\,{\rm kbar}$.
Thus on the tail's side, a redistribution of electronic properties may occur. 
As its weight is low (1\%), it is not obvious to focus and zoom on this weak contribution. 
The challenge is now to detect via a direct local probe such as scanning tunneling microscopy if such exotic effects exist. 
Previous NMR experiments have shown that a residual AF component survives in parallel with a main signal assigned to the intrinsic HO.\cite{Matsuda2001}
It would be interesting to see if this can be observed systematically in high quality single crystals too. In general, a difficulty is that in heavy fermion compounds, the microscopic observations of defects
have been never clear as has been shown by detailed studies on UPt$_3$.\cite{1UPt3,2UPt3}
In the well known Ce-115 family, mixing of different phases has been observed. E.g.~for the antiferromagnet CeRhIn$_5$, persistence of SC and commensurate magnetically ordered phase is observed even at zero pressure while its is expected to occur only at high pressures \cite{1CeRhIn5,2CeRhIn5}. In contrast, in CeIrIn$_5$ bulk superconductivity occurs at $T \sim 0.4$~K while the superconducting resistive transition appears already at $T\sim 1.2$~K.\cite{Petrovic2001} 

The longitudinal magnetoresistivity experiments show that 
the Fermi liquid regime will be achieved only below $T_{\rm SC}$.
The fact that RRR dependences of $T_{\rm SC}$ and $T_{\rm H0}$ have quite similar absolute shifts 
confirms the role of itinerancy of the 5\textit{f} electrons of uranium atoms in both transitions.
The unexpected phenomena are that increasing RRR does not lead to an unique SC transition,
while below $T_{\rm SC}$, so far as RRR is larger than 30, excellent agreement is observed
in the temperature dependence of $C/T$ with extrapolated electronic value of $\gamma = (C/T)_{T\to 0}$
near zero.

\begin{table*}[htb]
\caption{  Critical pressures for the transition from HO to AF state $P_x$ and the midpoint of the SC transition $P_{\rm SC}^{\rho, {\rm mid}}$ and the maximal pressure of a complete SC transition $P_{\rm SC}^{\rho=0}$ (defined by different experimental probes) as reported in literature. RRR gives the residual resistivity ratio.}
\label{table4}
\begin{center}
\begin{tabular}{l l l cccccccc  } \hline 
Year &		Reference			&pressure medium	& RRR 		& 	$P_x$  	& 	$P_{\rm SC}^{\rho, {\rm mid}}$ & $P_{\rm SC}^{\rho=0}$ & $P_{\rm SC}^{C}$ &$P_{\rm SC}^{\chi}$  \\ \hline \hline
1987	&  McElfresh\cite{McE} 	& 1:1 isoamyl/n-pentane	& $\sim 20$ &	 		& 	1.6		& $< 1$ \\
1993	& Schmidt\cite{1993Shumidt}	& 1:1 isoamyl/n-pentane	&$\sim 35$ &			& 	1.2	& $<1 $	\\
2003	& Motoyama\cite{2003Motoyama}& 1:1 FC70 / FC77  & 20 - 30 & $   0.3-0.7$	&	&	&& \\
2005	& Tenya\cite{2005Tenya} 	&		&	&	&&&	&$<$1.5  \\
2006	& Sato\cite{2006Sato}		&   &	30 &	0.35	& 	&& &	0.45		\\
2007	& Jeffries\cite{2007Jeffries,2008Jeffries} & 1:1 isoamyl/n-pentane & $\sim 7$  & 	 &	$ \sim 1.35$ & 0.8 & \\
2008	& Amitsuka\cite{2008Amitsuka}&  1:1 FC70 / FC77& & 0.7		&	 	&		& 	&	0.7		\\
2008	& Hassinger\cite{2008Hassinger} &  Argon	&$\sim 25 $ &  0.47		& 		$\sim 2 $ & 1.8 & 0.5	\\
2008	& Motoyama\cite{2008Motoyama} &  Daphne 7373	&  &  0.45		& 		 	\\
2010	& Niklowitz\cite{2010Niklowitz}& 1:1 FC70 / FC77 &$\sim 20$  & 0.45	&	&	&\\
2010	& Butch\cite{2010Butch} & Helium	&  $<10$ &	0.8		& 	0.8			\\
2011	& Hassinger\cite{2011Hassinger} & Daphne 7373& $\sim 160 $ &  	& 		$  \approx 1.3 $  & $ \approx 1.1 $\\
2011	& Tateiwa\cite{2011Tateiwa,2011Tateiwa2} & Daphne 7474	& $\sim 300$	&  	& 	$\approx 1.5$	& $\approx 1.3$	\\
\hline
\end{tabular}
\end{center}
\end{table*}

Finally, we want to stress that the high pressure phase diagram of URu$_2$Si$_2$ is now very well established and a broad consensus has been achieved qualitatively. 
In Tab.~\ref{table4} we summarize the critical pressures $P_x$ and $P_{SC}$ as reported in the literature. 
Of course the table is not complete, but it gives a good review of different experiments. 
It is widely established that the transition from the HO state to the AF state is of first order and bulk superconductivity is suppressed when long range AF order appears at high pressure. 
However, as can be seen in Tab.~\ref{table4}, the exact determination of the phase transition lines depends on (i) the sample quality, but also on (ii) the hydrostatic pressure conditions.  
Systematic studies of the influence of the different pressure conditions have been reported in refs.~\citen{2009Butch,2010Butch}, 
and the claim is that under hydrostatic conditions the HO -- AF transition appears at $P_x \approx 0.8$~GPa. 
However, the investigated samples have a rather bad RRR $<$ 10.  There is no detailed study of the influence of sample quality on the high pressure phase diagram. 
This is mainly due to the fact that very high quality single crystals (at least regarding the RRR) are small in size and mainly from the outer part of the single crystals with a free surface and experimentally, 
the phase line HO -- AF is rather difficult to draw from macroscopic measurements. 
Microscopic experiments such as neutron scattering ask for large single crystals.  
Furthermore, the usually used pressure medium for neutron scattering experiments (1:1 mixture of fluorinert FC70 / FC77) lacks hydrostaticity. 
However, in ref. \citen{2008Amitsuka} neutron scattering measurements
with two different pressure transmitting media reveal that with better hydrostaticity $P_x$ becomes smaller.

Pressure measurements indicate that at $P_x \approx 0.5\,{\rm GPa}$ the HO switches to AF. 
Uniaxial stress experiments emphasizes recently that the change will occur at $\sigma_a = 0.35\,{\rm GPa}$ for a stress applied long the $(1,0,0)$ basal plane.\cite{Yokoyama2005}
Lately, microscopic and macroscopic measurements show that depending on the pressure conditions 
$P_x$ can vary from $0.47\,{\rm GPa}$ to $0.8\,{\rm GPa}$ (see Tab.~\ref{table4}).
Furthermore while there is now no doubt that bulk superconductivity disappears at $P_x$, 
it is observed that zero resistivity can be obtained up to $2P_x$ for example $P_{\rm SC}^\rho = 1.8\,{\rm GPa}$
instead of $P_x\sim 0.47\,{\rm GPa}$ by specific heat.
This difference supports the previous picture of internal pressure
or uniaxial stress distribution in the crystal.

The crystal quality of the data presented in Tab.~\ref{table4} is quite different and is another origin of the deviation of $P_x$. For example the crystal studied in ref.~\citen{2010Butch} has an RRR $\approx 7$, the resistivity at 10~K is near 40~$\mu \Omega$cm while it is less that 10~$\mu \Omega$cm in high quality single crystals. Compared to the best crystals a negative shift of $T_{\rm HO}$ of $0.2$~K appears in such lower quality crystals. Assuming a $\frac{\partial T_{\rm HO}}{\partial P} \sim 1.3$~K/GPa, the shift of $T_{\rm HO}$ can be viewed as a negative pressure of 0.2~GPa which may be the origin of the overestimation of $P_x$ by 0.2~GPa. 
In addition impurities can wipe out the feedback with lattice deformations. This may lead to another source of the $P_x$ displacement.  
In ideal conditions, $P_x$ seems to be near 0.5~GPa (see Tab.~\ref{table4}). 
It is amazing to observe that the convergence of $P_x$ to $P_c^\rho$ seems to occur for the low RRR case as if impurity scattering wipes out nanostructure phenomena. 
A sound determination of $P_c^\rho$ as indicated in Tab.~\ref{table4} is rather difficult as the resistive transition can get very broad under high pressure, 
or only a partial superconducting transition appears at highest pressures.\cite{2011Tateiwa} 
E.g.~in ref.~\citen{2008Jeffries} the superconducting transition at $P_c^\rho$ is almost 1~K and the onset of the superconducting transition vanishes only close to $P=2.8$~GPa. However, 
bulk superconductivity has been shown to collapse at $P_x$.\cite{2006Sato, 2007Amitsuka, 2008Hassinger}

Thus it is worthwhile to emphasize that the effects of non-hydrostaticity and crystal purity have drastic consequences on the boundary limit of URu$_2$Si$_2$. 
An underestimated phenomena is that the effects of either internal residual strain or 
external strain produced for example by fixing a tiny crystal on a sample holder will be quite different depending on the crystallographic orientation of the strain. 
For (0,0,1), the tetragonal symmetry will be preserved; for (1,0,0) and (1,1,0) 
it will drive to orthorhombic symmetry with strong differences between (1,0,0) and 
(1,1,0) axis as observed in recent magnetostriction experiments\cite{Hardy2011} and stressed in new theoretical developments \cite{Kusunose2011}. 
Full understanding of basal plane anisotropy \cite{Okazaki2011} requires careful cross checking.

Finally, let us notice that the link between a determination of the energy gap $\Delta \sim 7$~meV (70~K) in Tabs.~\ref{tab} and~\ref{table2} 
and recent microscopic measurements is not obvious. 
Below $T_0$ inelastic neutron scattering experiments show clear excitations at 1.7~meV and 4~meV for the wave vectors $Q_0 = (1, 0, 0)$ and $Q_1 = (0.6, 0, 0)$ \cite{1987Broholm, Bourdarot} 
while scanning tunneling microscopy experiments points out a bias asymmetric energy gap \cite{Aynajian2010}, 
recent laser angular resolved photo-emission spectroscopy indicates the formation of a gap at the Fermi surface energy around 2.9~meV \cite{Yoshida2010}, 
and a previous one detected a narrow peak at binding energy $E_B \sim -7$~meV.\cite{Santander2009}

A recent review on the unresolved case of URu$_2$Si$_2$ can be found in ref.~\citen{Mydosh2011}. 
Here we have concentrated on the link between material issues and physical properties.

\section{Conclusions}

Resistivity and specific heat experiments on URu$_2$Si$_2$ samples with very different sample quality have been reported in this article 
and the attempt has been made to associate the superconducting transition temperature $T_{\rm SC}$ 
and the HO transition temperature $T_{\rm HO}$ to the sample quality which is measured by the residual resistivity ratio RRR. 
We have shown that for RRR $>$ 50 both, $T_{\rm SC}$ and $T_{\rm HO}$, are almost independent of the RRR. 
However, even for samples with very high values of the RRR, 
it is difficult to obtain a single superconducting transition and $T_{\rm SC}$ obtained in resistivity experiments is always higher than in specific heat.  
In resistivity experiments no $T^2$ dependence is observed directly in high quality crystals indicating that the Fermi liquid regime is restricted to lower temperatures. 
In URu$_2$Si$_2$ the competition between superconductivity and antiferromagnetism are obvious and small intrinsic pressure gradients due to imperfections seems sufficient to induce a tiny volume fraction of SC in the AF state and vice versa. 
The ``residual" phenomena is believed to be extrinsic (tiny ordered moment -- extra superconducting network).  
But may be, it is the signature of new intrinsic features of a specific ``nanostructure" arrangement. 
The interesting paradox seems to be that these ''heterogeneities" are also observed in the best crystals with high values of the RRR. 
To obtain the ideal material of URu$_2$Si$_2$ seems to be a challenge for the future as does the search for contrasting effects in local probe spectroscopy.

\section*{Acknowledgements}  
The present work was financially supported by Grants-in-Aid for Young Scientists (B)(22740241), Scientific Research C (21540373, 22540378),
 Scientific Research S (20224015), Scientific Research on Innovative Areas ``Heavy Electrons" (20102002, 23102726) and 
 and Osaka University Global COE program (G10) 
from the Ministry of Education, Culture, Sports, Science and Technology (MEXT) and Japan Society of the Promotion of Science (JSPS). 
This work was also supported by ERC starting grant (NewHeavyFermion) and French ANR project (CORMAT, SINUS).
We thank S. Kambe for helpful discussion.

\bibliography{apssamp}

\end{document}